\documentstyle[twocolumn,aps,prl,epsf]{revtex}

\def\a{\alpha}		 		   \def\g{\gamma}	
		 \def\e{\epsilon}                 
\def\w{\omega}		 \def\l{\lambda}	   		
		 \def\D{\Delta}	         \def\W{\Omega}	
\def\G{\Gamma}	
\def\vk{{\bf k}}         \def\vr{{\bf r}}          
\def\H{{\cal H}}         \def\cG{{\cal G}}         \def\cF{{\cal F}}  
\def\Gc{{\cal G}_c}      \def\Gs{{\cal G}_s}       
\def\Dp{\D_+}	         \def\Dn{\D_-}	   
       \def\eF{\e_{\rm F}} 	   
\def\/{\over}            \def\kF{k_{\rm F}}	   \def\pF{k_{\rm F}}
\def\[{\left[}           \def\]{\right]}	   
\def\({\left(}           \def\){\right)}	   
\def\<{\langle}          \def\>{\rangle}

\begin{document}
  \twocolumn[\hsize\textwidth\columnwidth\hsize\csname
  @twocolumnfalse\endcsname
\draft

\title{Crossed Andreev reflection
in a $d$-wave superconductor \\with two quantum point contacts}

\author{S. Takahashi, T. Yamashita, and S. Maekawa}

\address{
Institute for Materials Research, Tohoku University, Sendai 980-8577, Japan}

\date{December 27, 2003}
\maketitle
\widetext

\begin{abstract}
We theoretically study the crossed Andreev reflection in a 
hybrid nanostructure which comprises a $d$-wave superconductor
and two normal-metal quantum wires.  
When the superconductor of the (110) oriented surface is
in contact with the wires parallel and placed close to each other,
the Andreev bound state is formed by the crossed Andreev reflection.
When the contact barrier potential is sufficiently large,
two sharp peaks appear in the conductance well below the gap
structure, which originate from the bonding and antibonding Andreev
bound states.  We propose that these Andreev bound states form a two-level
quantum system (qubit).
\end{abstract}

\pacs{PACS numbers: 74.45.+c,74.50.+r,81.07.Lk,73.40.Gk}

  \vskip1.0pc]
\narrowtext


Quantum transport in nanostructures is of current interest in both
theoretical and experimental studies.
In a quantum wire, its width is so narrow that electron waves are
strongly confined in the transverse direction and their transverse
momentums are quantized.  
In a ballistic conduction of electrons through a quantum point contact
with width comparable to the Fermi wavelength, a steplike structure
with step of $2e^2/h$ appears in the conductance as a function of
Fermi energy or width \cite{wees,wharam,szafer}.

One of the fundamental consequences of superconductivity is the
Andreev reflection at the interface of a normal-metal and a
superconductor (SC) \cite{andreev,BTK}.  This phenomenon corresponds
to an incoming electron from the normal side being reflected as a hole,
thereby adding a Cooper pair in the superconducting condensate.
In a tunnel junction of a normal-metal and a (110) oriented
$d$-wave SC, the zero bias conductance peak appears
due to the formation of the Andreev bound state
at the interface \cite{tanaka,kashiwaya}.
However, when a single quantum wire of a single conducting channel is
in contact with $d$-wave SC of the (110) oriented surface, the Andreev
reflection is completely suppressed due to the quantum mechanical
diffraction of electron waves at the narrow opening \cite{takagaki}.

A basic question arises what happens if two quantum wires are
in contact to the (110) oriented $d$-wave SC (see Fig.~1).
When an electron is injected into SC from one of the wires, there is
a possibility that the Andreev hole is reflected back into another
wire due to the non-local effect called the {\it crossed} Andreev
reflection (CAR) \cite{bayers,deutcher},
thus providing an ideal system to study CAR.

In this Letter, we explore quantum-interference effects due to the
crossed Andreev reflection in a $d$-wave SC with two quantum wires
that are parallel and placed close to each other.
It is shown that the resonance peak in the conductance is split into
two sharp peaks at low energies when the barrier potential of the
contact is sufficiently large.  The lower and higher
energy peaks correspond to the bonding and the antibonding Andreev
bound states, respectively.   This suggests that these Andreev levels
form a two-level quantum system (qubit), whose population can be
controlled by application of bias voltage and/or electromagnetic field,
exhibiting a coherent oscillation of the Andreev qubit (Rabi oscillation).

\begin{figure}[thb]					
  \epsfxsize=0.72\columnwidth				
  \centerline{\hbox{\epsffile{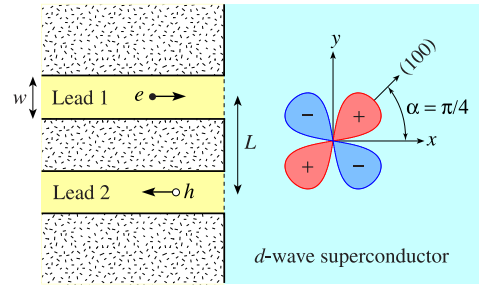}}}	
  \vskip 0.40cm
  \caption{ \small					
Schematic diagram of two-leg normal-conducting quantum
wires in contact with a $d$-wave superconductor.
  }   \label{fig1}					
\end{figure}

We examine the quantum transport in a hybrid nanostructure of
a $d$-wave SC and two normal-conducting quantum wires.
Figure~\ref{fig1} shows a model structure in the $x$-$y$
plane with a two-dimensional (2D) $d$-wave SC occupying
the left half space, and two quantum wires of width $w$, lead 1 and lead 2,
which are parallel along $x$ and connected to SC
at $y=\pm L/2$.
The wave functions of electron and hole like quasiparticles with
excitation energy $E$ in the electrodes are determined by the
Bogoliubov-de Gennes equation
 \begin{eqnarray}
   \left( 
     \matrix{ {\H_0}   &   \D(\vk,\vr)  \cr \D^*(\vk,\vr) &  -{\H_0}  }
     \right)  \left( \matrix{ u_\vk(\vr) \cr v_\vk(\vr) }  \right)
    =   E     \left( \matrix{ u_\vk(\vr) \cr v_\vk(\vr) }  \right) ,
 \end{eqnarray}
where $\H_0=-\hbar^2\nabla^2/2m-\eF$ is the single-particle
Hamiltonian with the Fermi energy  $\eF=\hbar^2\kF^2/2m$.  
For simplicity, the Fermi wave number $\kF$ and the effective mass
$m$ are common for all electrodes, and the amplitude of the gap function
$\D(\vk,\vr)$ is uniform in SC and vanishes
 in the wires.  In an anisotropic SC, the pair potential $\D(\vk,\vr)$
is a function of wave vector $\vk$, and its value is
defined on the Fermi surface in the direction of $\vk$.
In leads 1 and 2 with infinite wall boundaries,
the wave

\vfill
\noindent
 function of an electron propagating along the $x$ axis
with wave number $k$ in the $n$th electron subband is given by
the product of the plane wave $e^{ikx}$ and the $n$th transverse wave,
$\chi_n(y - L/2)$ in lead 1 and $\chi_n(y + L/2)$ in lead 2,
where $\chi_n(y)=(2/w)^{1/2} \sin\[(n\pi/w)\(y+w/2\)\]$, and has
the energy $E=(\hbar^2/2m)[k^2+(n\pi/w)^2]-\eF$. 
In the following, we restrict our calculation to the case where
only the lowest subband ($n=1$) is occupied by electrons (or holes)
in the wires, which is realized for those wires satisfying
$\pi < \kF w <2\pi$.  
When an electron with energy $E$ and wave number
$k_{1} = [2mE+\kF^2 - (\pi/w)^2]^{1/2}$ is incident from lead 1
into SC, the wave functions in leads 1 and 2 are given by
$\Psi_{1}(x,y)=\varphi_{1}(x)\chi_1(y-L/2)$ and
$\Psi_{2}(x,y)=\varphi_{2}(x)\chi_1(y+L/2)$, where
\begin{eqnarray}
  \varphi_{1} &=&
      \( \matrix{1 \cr 0} \) e^{ ik_{1}x}
       + r_{\rm 11}^{ee}\( \matrix{1 \cr 0} \) e^{-ik_{1}x} 
       + r_{\rm 11}^{eh}\( \matrix{0 \cr 1} \) e^{ ik_{1}x}  ,  
  \label{eq:fN1} \\
  \varphi_{2} &=&
     r_{\rm 12}^{ee}\( \matrix{1 \cr 0} \) e^{-ik_{1}x} 
       + r_{\rm 12}^{eh}\( \matrix{0 \cr 1} \) e^{ ik_{1}x} .
  \label{eq:fN2} 
\end{eqnarray}
Here, $r_{11}^{ee}$ and $r_{11}^{eh}$ are the amplitudes of the
normal reflection (NR) and Andreev reflection (AR),
respectively, while $r_{\rm 12}^{ee}$ and $r_{\rm 12}^{eh}$ are those of
the \textit{crossed} normal reflection (CNR) and the \textit{crossed}
Andreev reflection (CAR), respectively.   
A similar treatment is made for an incident electron from lead 2.
Since $E \alt \D \ll \eF$, we put $k_{1} \approx \kF [ 1 - (\pi/\kF w)^2]^{1/2}$
in the following.

We employ the Blonder-Tinkham-Klapwijk (BTK) approach \cite{BTK}
to calculate the conductance of the structure.
To carry out the calculation analytically, we make the Andreev
approximation that neglects all the evanescent modes \cite{approx},
and put the wave function of SC in the form
\begin{eqnarray}
  \Psi_{s}(x,y) &=&
    \int_{-\kF}^{\kF}
     t_s^{ee}(p_y) \( \matrix{ 1 \cr  \G_+ } \)
     e^{i\sqrt{\pF^2-p_y^2}x} e^{ip_yy}  dp_y
       \cr &+&
    \int_{-\kF}^{\kF}
     t_s^{eh}(p_y) \( \matrix{ \G_- \cr 1 } \)
     e^{-i\sqrt{\pF^2-p_y^2}x} e^{ip_yy}  dp_y,
   \label{eq:PsiS}
\end{eqnarray}
where $\G_+={\D_+^* / (E+\W_+)}$ and $\G_-={\D_- / (E+\W_-)}$
with $\W_\pm=\sqrt{E^2-|\D_\pm|^2}$ \cite{tanaka,kashiwaya,takagaki}.
The first and second terms in Eq.~(\ref{eq:PsiS}) are the transmitted
QP waves on the electron-like and hole-like branches, respectively.
In SC of $d$-wave symmetry, QPs in 
different branches feel different pair potentials,
$\Dp = \D_0\cos2(\theta-\a)$ and $\Dn = \D_0\cos2(\theta+\a)$,
where $\a$ is the angle between the (100) axis of SC and
the normal to the interface (see Fig.~\ref{fig1}), and 
$\theta=\sin^{-1}(p_y/\pF)$ is the propagation angle relative
to the $x$ axis.

In the following, we focus on the (110) oriented surface of SC,
i.e.,  $\alpha=\pi/4$, as shown in Fig.~\ref{fig1}.  
In this case, QPs in the electron-like and hole-like branches
move in the pair potentials of opposite sign, $\D_\pm=\pm\D$,
where 
{$\D = 2\D_0(p_y/\pF)\sqrt{1-(p_y/\pF)^2}$}, and
$\W_\pm=\W=\sqrt{E^2-\D^2}$.
The barrier potential at the interface between the wires and SC
is taken into account by the $\delta$-function-type potential
with amplitude $(\hbar^2\pF/2m)Z$, $Z$ being a dimensionless
parameter \cite{BTK}.
The boundary conditions for the wave functions at the interfaces
are $\Psi_{s}(0,y)=\Psi_{i}(0,y)$
and $[\partial_x{\Psi_{s}(x,y)}-\partial_x{\Psi_{i}(x,y)}]_{x=0}=\pF Z\Psi_{i}(0,y)$
 ($i=1,2$) appropriate for the $\delta(x)$ potential.
The matching technique to the boundary conditions \cite{szafer}
yields the reflection coefficients
\begin{eqnarray}
  r_{11}^{ee} &=&  -1  + {\bar{k}_1\/{\cal D}_+}({\bar{k}}_1+\cF-\Gc-iZ)  \cr
      & & \ \ \ \ \    + {\bar{k}_1\/{\cal D}_-}({\bar{k}}_1+\cF+\Gc-iZ) ,  
    \label{eq:a1}  \\
  r_{11}^{eh} &=& 
       -{\bar{k}}_1{\cG}_s \( {1\/ {\cal D}_-} - {1\/ {\cal D}_+}\),
   \label{eq:b1}  \\
  r_{12}^{ee} &=&
     - {2{\bar{k}}_1\Gc\/ {\cal D}_+ {\cal D}_-} \[ ({\bar{k}}_1+\cF-iZ)^2-(\Gc^2+\Gs^2)\] 
         ,
   \label{eq:a2}  \\
  r_{12}^{eh} &=&
     - {\bar{k}}_1{\cG}_s \( {1\/ {\cal D}_-} + {1\/ {\cal D}_+}\),    
   \label{eq:b2}
\end{eqnarray}
with $\bar{k}_1=k_{1}/\kF$,
${\cal D}_\pm=({\bar{k}}_1+\cF)^2-(\Gc \pm iZ)^2-\Gs^2 $,
and
\begin{eqnarray}
  \cF &=& 
      \int_{-\kF}^{\kF}   {d{p}\/ 2\pi} 
      {\W\/ E}
     \sqrt{1- (p/\kF)^2} \varphi^2(p),
      \label{eq:F1}  \\
  \Gc &=& 
      \int_{-\kF}^{\kF} {d{p}\/ 2\pi} 
      {\W\/ E}
     \sqrt{1- (p/\kF)^2} \varphi^2(p)
     \cos(pL) ,
     \label{eq:G1}  \\
  \Gs &=& 
     i \int_{-\kF}^{\kF} {d{p}\/2\pi} {\D\/E}
     \sqrt{1- (p/\kF)^2} \varphi^2(p)
     \sin(pL) ,
     \label{eq:G2}
\end{eqnarray}
where $\varphi(p)=\<{p}|\chi_1\>$ is the overlap integral of $\chi_1(y)$
and $e^{i{p}y}$:
\begin{eqnarray}
  \varphi(p)  = \sqrt{8 w/\pi^2} {\cos(pw/2)/\[ 1-(pw/\pi)^2 \]}.
\end{eqnarray}
Note that $\cF$ in Eq.~(\ref{eq:F1}) represents the local coupling and
is independent of distance $L$, while $\Gc$ and $\Gs$ represent the non-local
coupling and are dependent on $L$.
In the limit of $L\rightarrow\infty$, where the two contacts are
independent ($\Gc=\Gs=0$), one has
$r_{11}^{ee}=(k_{1}-\cF-iZ)/(k_{1}+\cF+iZ)$, 
$r_{11}^{eh}=r_{12}^{ee}=r_{12}^{eh}=0$,
recovering the complete suppression of AR in a single quantum wire
with a single transverse mode \cite{takagaki}.
The reflection coefficients $r_{22}^{ee}$, $r_{22}^{eh}$, $r_{21}^{ee}$,
and $r_{21}^{eh}$ for an incident electron from lead 2 are 
obtained from those in Eqs.~(\ref{eq:a1})-(\ref{eq:b2}) 
by the replacement $1 \leftrightarrow 2$ and $L \rightarrow -L$.
It follows from Eqs.~(\ref{eq:b1})-(\ref{eq:b2}) that, when there is no
barrier potential ($Z=0$) at the interface, AR is absent ($r_{11}^{eh}=0$),
whereas CNR and CAR are finite with the ratio $r_{12}^{ee}/r_{12}^{eh}=\Gc/\Gs$.

\begin{figure}[thb]					
  \epsfxsize=0.92\columnwidth				
  \centerline{\hbox{\epsffile{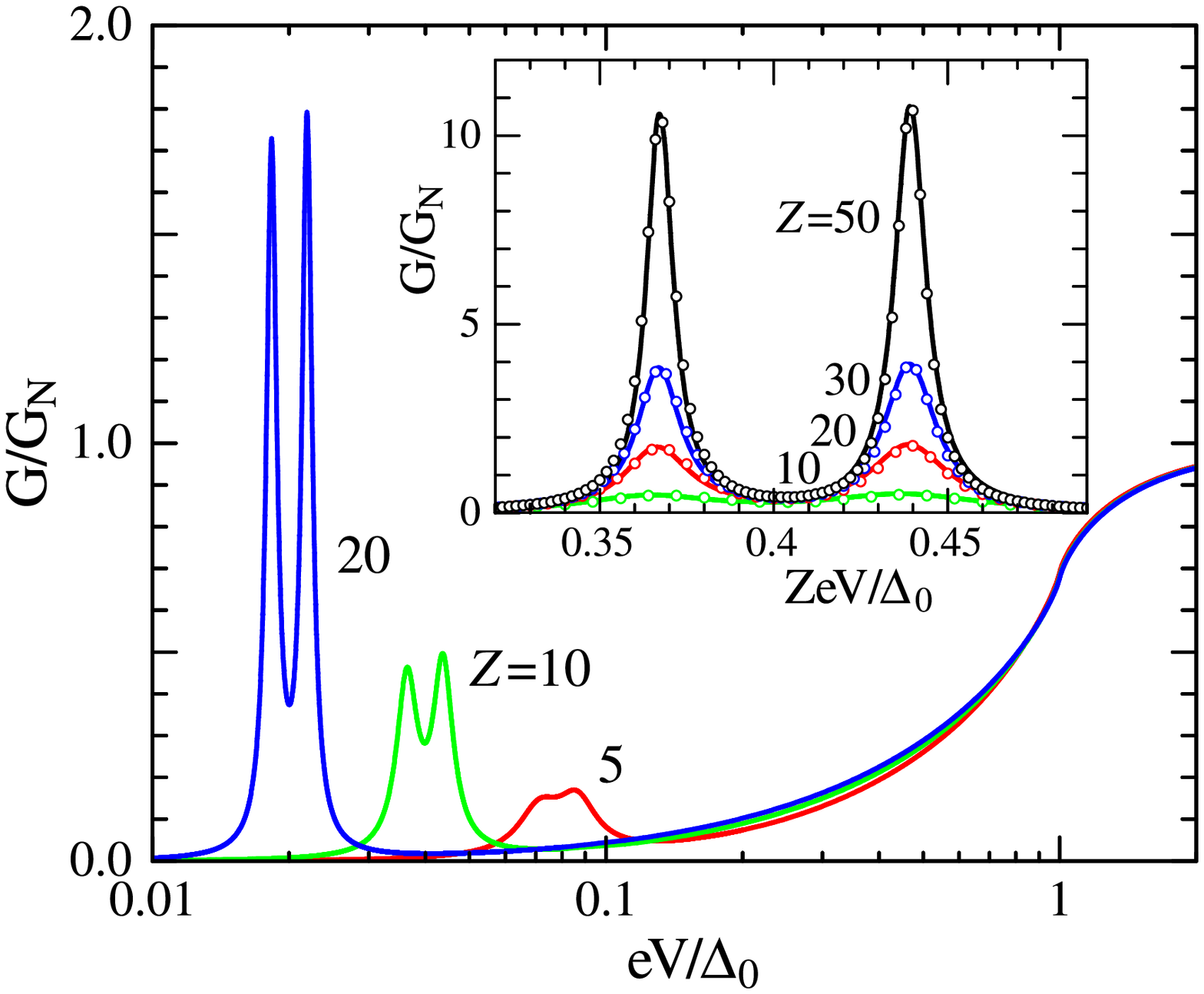}}}	
  \vskip 0.40cm
  \caption{						
Normalized conductance $G/G_N$ as a function of bias voltage $V$
for different values of interfacial barrier parameter $Z$.
Inset shows $G/G_N$ vs normalized voltage $ZV$.
The circles in the inset are calculated from Eq.~(\ref{eq:GsGn}).
  }   \label{fig2}					
\end{figure}

When bias voltage $V$ is applied to the two leads, 
the conductance $G$ at zero temperature ($T=0$)
is given by
\begin{eqnarray}
  G={4e^2\/h} (1-|r_{11}^{ee}|^2-|r_{12}^{ee}|^2
                       +|r_{11}^{eh}|^2+|r_{12}^{eh}|^2)_{E=eV},
  \label{eq:Gs}
\end{eqnarray}
where $h$ is the Planck constant.
Figure~\ref{fig2} shows the conductance $G$ vs $V$ for
$\kF w=4$, $\kF L=8$, and different values of $Z$. \
The conductance is normalized by the normal state value $G_N=G(\D=0)$, which takes 
$G_N \sim 16(e^2/h)(k_{1}/\kF)\cF_N/Z^2$ ($Z \gg 1$) with $\cF_N=\cF(\D=0)$.

\noindent
For a low barrier potential of small $Z$, the conductance decreases
monotonically with decreasing $eV$ below $\D_0$.
As the barrier potential becomes higher, a peak structure appears well
below $\D_0$ and shifts towards lower $eV$, developing the double peak
structure with increasing $Z$.
If the conductance is plotted as a function of normalized voltage
$ZV$ as shown in the inset, the conductance peaks fall into the same
position, indicating that the resonance peak positions are
scaled by $1/Z$.

Let us examine the origin of the double peak structure in the conductance
for the tunneling case ($Z \gg 1$).  Since $\W \approx i|\D|$ for $E \ll \D_0$,
$\cF$, $\Gc$, and $\Gs$ in Eqs.~(\ref{eq:F1})-(\ref{eq:G2})
have the forms: 
$\cF \approx i f \D_0/E$, $\Gc \approx i g_c\D_0/E$, and
$\Gs = i g_s\D_0/E$,
where 
${f} =(E{\cal F}/i\D_0)_{E\rightarrow 0}$,
${g_c} =(E\Gc/i\D_0)_{E\rightarrow 0}$, and
${g_s} =(E\Gs/i\D_0)$ are energy independent quantities, 
so that the Andreev reflection coefficients are calculated in the
resonance forms
\begin{eqnarray}  
    r_{11}^{eh} &\approx&
      - { ig\g \/ E-{E_-} + i\g } 
      + { ig\g \/ E-{E_+} + i\g },
  \label{eq:A-b1}  \\
    r_{12}^{eh} &\approx&
      -  { ig\g \/ E-{E_-} + i\g } 
      - { ig\g \/ E-{E_+} + i\g }, 
 \label{eq:A-b2}
\end{eqnarray}
where
$g=g_s/2f$, $E_\pm=  (f \pm |g_c|+g_s^2/2f)\D_0/Z$, and
$\g=(k_{1}/\kF)f\D_0/Z^2$.   
The other coefficients are $r_{11}^{ee} \approx -1-gr_{12}^{eh}$
 and $r_{\rm 12}^{ee} \sim O(g^2)$.
The resonance energies $E_\pm$ and intensity $g$ depend strongly on
lead separation $L$, exhibiting damped oscillations with the period
of the Fermi wave length $\l_{\rm F}=2\pi/\kF$.
We note that the resonance positions $E_\pm$ and their separation
$E_+-E_-$ are scaled by $1/Z$, while the line width $\g$
is scaled by $1/Z^2$.   For $Z \gg 1$, the line width of the peaks
is much smaller than the separation, and therefore a well-separated
two-peak structure is formed as shown in the inset of Fig.~\ref{fig2}.
In this case, the conductance $G$ is written as the sum of two Lorentzians
\begin{eqnarray}  
   {G \/ G_N}  \approx  
       { {\cal I} \g/\pi \/ (eV-{E_+})^2 + \g^2} 
     + { {\cal I} \g/\pi \/ (eV-{E_-})^2 + \g^2}, 
  \label{eq:GsGn}
\end{eqnarray}
where ${\cal I}=(\pi g^2f/4\cF_N)$.
The  simple formula (\ref{eq:GsGn}) reproduces the numerical
result in the inset of Fig.~\ref{fig2}, if the calculated values ($f=0.405$, $g_c=-0.036$,
$g_s=-0.0345$, and $\cF_N=0.68$) are used in Eq.~(\ref{eq:GsGn}).
Note that the peak height of $G/G_N$ increases in proportion to $Z^2$.

\begin{figure}[thb]					
  \epsfxsize=0.87\columnwidth				
  \centerline{\hbox{\epsffile{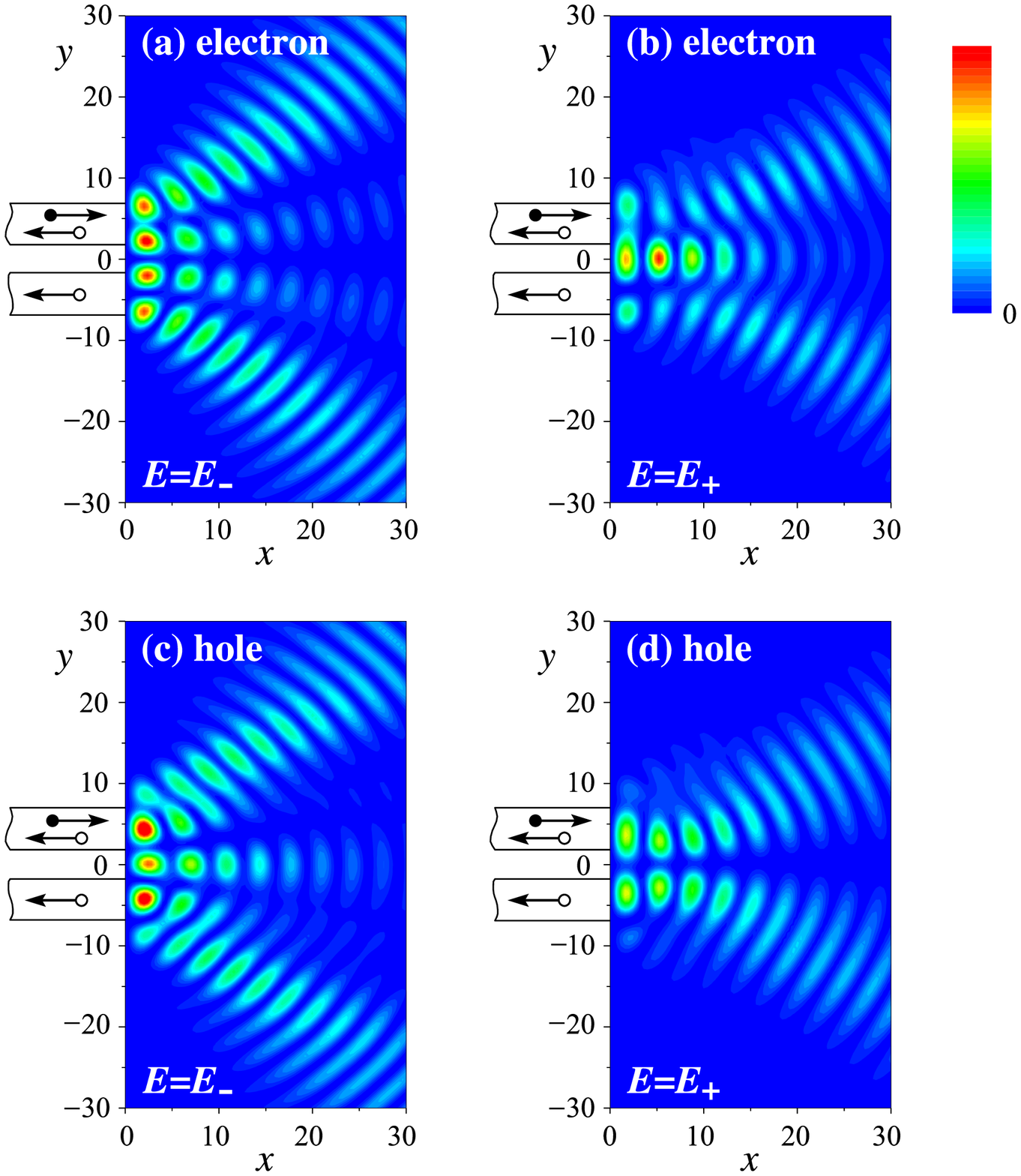}}}	
  \vskip 0.40cm
  \caption{						
Mapping of Andreev bound states on the $xy$ plane,
when an electron is incident from lead 1 to a $d$-wave SC.
(a) and (b) are the absolute square of electron
wave functions $|\Psi^e_{s}|^2$ at energies $E_-$ and $E_+$;
(c) and (d) are that of hole wave functions
$|\Psi^h_{s}|^2$ at $E_-$ and $E_+$.
  }   \label{fig4}					
\end{figure}

To elucidate the formation of the Andreev bound states, we calculate
the QP wave function $\Psi_{s}(x,y)=$ $^t(\Psi^e_{s},\Psi^h_{s})$ in SC,
where $\Psi^e_{s}$ and $\Psi^h_{s}$ are
the electron and hole wave functions, respectively.
Figure~\ref{fig4} shows the mapping of the absolute squares, $|\Psi^e_{s}|^2$
and $|\Psi^h_{s}|^2$, at the resonance energies on the $xy$ plane
 for $\kF w=5$, $\kF L=8$, and $Z=50$,
when an electron is incident from lead 1.  It is clearly seen that
the QP wave functions are strongly localized with very large peaks
(red color) near the contacts due to the formation of the Andreev bound states,
and that the decaying QP waves are traced to the (010) and (0${\bar 1}$0)
directions along which multiple reflections of electron and hole take place.
It is also seen that the formation of the localized states,
especially the relative position of the peaks with respect to the contacts,
is different between the electron and hole QPs, and between the resonance
energies $E_\pm$.
For $Z \gg 1$ and around $E_\pm$, the explicit form of $\Psi_{s}$ inside
SC is obtained as
\begin{eqnarray}
    \Psi_{s}
    \approx  { i\g \/ E-{E_-} + i\g } 
             \(\matrix{\psi_{-}^e \cr \psi_{-}^h}\)
           + { i\g \/ E-{E_+} + i\g } 
             \(\matrix{\psi_{+}^e \cr \psi_{+}^h}\),
   \nonumber
\end{eqnarray}
where, except very close to the interface ($x \approx 0$),
\begin{eqnarray}
    \(\matrix{\psi_{-}^e  \cr \psi_{-}^h}\) &=& {2g\/\pi}
       \int_{0}^{\kF}  {\D\/E} \varphi(p) \cos{pL\/2}
      \sin{qx} \(\matrix{\ \ \ \sin py \cr  -\cos py}\) {dp},
       \nonumber \\
   \(\matrix{\psi_{+}^e  \cr \psi_{+}^h}\) &=& {2g\/\pi}
       \int_{0}^{\kF}  {\D\/E} \varphi(p) \sin{pL\/2}
       \sin{qx} \(\matrix{\cos py \cr \sin py}\) {dp},
       \nonumber
\end{eqnarray}
with $q=\sqrt{\kF^2-p^2}$ and $\D = 2\D_0pq/\pF^2$.
At lower resonance energy $E_-$, $\Psi_{s}$ is dominated by the first term
whose electron (hole) wave function $\psi^e_{-}$ ($\psi^h_{-}$)
is an odd (even) function of $y$.   
At higher resonance energy $E_+$, $\Psi_{s}$ is dominated by the second term
whose electron (hole) wave function $\psi^e_{+}$ ($\psi^h_{+}$)
is an even (odd) function of $y$.
These results indicate that the electron (hole) wave functions at the lower
and higher bound states have different parity with respect to $y$.
It is noteworthy that the Andreev holes are reflected back into
leads 1 and 2 in phase at $E_-$ and out of phase at $E_+$
(cf. Eqs.~(\ref{eq:A-b1}) and (\ref{eq:A-b2})).


The Andreev bound states appeared in the present system have the following
implication.  When the interface barrier potential is very high ($Z \gg 1$),
the current flows through the system via the Andreev bound states with
very sharp energy levels at $E_\pm$.
When the bias voltage $V$ is set between ${V_-}={E_-}/e$ and ${V_+}={E_+}/e$
of the conductance peaks,
the lower bound state at $E_-$ is occupied by Andreev quasiparticles
while the higher-energy bound state at $E_+$ is empty;
the lower bound state is viewed as

\noindent
the ground state and the upper bound
state as the excited state, and therefore the system in this setup
forms a two-level quantum system (qubit).
For simplicity, we denote the ground state and excited state by
$|0\>$ and $|1\>$, respectively.
The Andreev bound states of electron (hole) QP at $E_-$ and $E_+$ have
dipole moments $\<1|y|0\>$
due to the different parity of the states with respect to $y$.
Therefore application of the electric field $eE_0\cos(\w t)$ oscillating 
along $y$ causes a significant Rabi oscillation with frequency
${\cal V}=eE_0\langle1|y|0\rangle/\hbar$ between the Andreev bound states,
if $\w$ is tuned to the resonance condition $\w=E_+-E_-$ and
the damping rate $\g$ of the states is substantially smaller than
$\hbar{\cal V}$.  The significant coherence oscillation requires a
smaller damping $\g$, which is achieved by inserting tunnel barriers
at the interfaces.
The controlled evolution between the two states $|0\rangle$ and
$|1\rangle$ is realized by applying resonant microwaves to the system.
The two level Andreev bound states in the proposed system provides
a new possibility of superconducting qubit ($d$-wave Andreev level qubit)
 \cite{zazunov}.

The realization of quantum two-level systems (qubit) is fundamental
issues in both physics and information technologies,
because qubits are basic elements for quantum computation \cite{book}.  
Recently, the coherent manipulation of the states in two-level
systems has been demonstrated in nanostructured superconducting circuits
\cite{nakamura,jj}.


In summary, we have theoretically studied the quantum-interference effects
caused by the crossed Andreev reflection
in a hybrid nanostructure which comprises a $d$-wave superconductor
and normal-conducting quantum wires.  
When a superconductor of the (110) oriented surface is in contact
with the two-leg quantum wires via tunnel barriers, the resonance bound
states are formed at low energies due to the crossed Andreev reflection.
As a consequence, two well-separated sharp peaks appear in the conductance
well below the superconducting gap structure.
The lower and higher conductance peaks correspond to the bonding
and the antibonding Andreev bound states 
whose wave functions have different parity.  
We propose that these Andreev bound states form a two-level
quantum system ($d$-wave Andreev level qubit), which can be controlled by
application of bias voltage and/or radiation of oscillating electric field.

This work was supported by a Grant-in-Aid for from MEXT,
NAREGI, and CREST, Japan.




\begin{thebibliography}{9}
\bibitem{approx}
This is is a good approximation when the width of the leads
satisfies the condition $1.5 \alt \kF w/\pi <2$ \cite{takagaki}.
\end{thebibliography}
\end{document}